




\magnification\magstep1
\baselineskip14pt
\vsize24.0truecm 
\hsize15.0truecm
\hoffset0.00truecm 

\def\hatt{\widehat}
\def\dell{\partial}
\def\tilda{\widetilde}
\def\eps{\varepsilon}
\def\half{\hbox{$1\over2$}}

\def\quart{\hbox{$1\over4$}}

\def\arr{\rightarrow}
\def\normal{{\cal N}}
\def\Var{{\rm Var}}
\def\E{{\rm E}}
\def\d{{\rm d}}

\def\mtrix{\pmatrix} 
\def\subsection{\medskip}

\font\bigbf=cmbx12 

\font\csc=cmcsc10
\font\smallrm=cmr8 

\def\today{\number\day \space \ifcase\month\or
January\or February\or March\or April\or May\or June\or 
July\or August\or September\or October\or November\or December\fi  
\space \number\year}

\def\ref#1{{\noindent\hangafter=1\hangindent=20pt
  #1\smallskip}}          

\def\hskipdistanceleft{\hskip-3.8pt}
\def\hskipdistanceright{\hskip-2.0pt}
\footline={{
\ifodd\count0
        {\hskipdistanceleft\quotationone\phantom{\today}
                \hfil{\rm\the\pageno}\hfil
         \phantom{\quotationone}{\smallrm\today}\hskipdistanceright}
        \else 
        {\hskipdistanceleft\quotationtwo\phantom{\today}
                \hfil{\rm\the\pageno}\hfil
         \phantom{\quotationtwo}{\smallrm\today}\hskipdistanceright}
        \fi}}
         

\def\today{November 1995}

\def\quotationone{\smallrm Hjort and Jones}
\def\quotationtwo{\smallrm Locally parametric densities} 

\def\sumin{\sum_{i=1}^n}
\def\bfx{{\bf x}}
\def\bft{{\bf t}}
\def\bfX{{\bf X}}
\def\frac#1#2{\hbox{${#1\over #2}$}}
\def\ident{\equiv}

\centerline{\bigbf Locally parametric nonparametric density estimation}

\smallskip
\centerline{\bf N.L.~Hjort and M.C.~Jones} 

\smallskip
\centerline{\bf University of Oslo and The Open University}

{{\medskip\narrower\noindent\baselineskip11.5pt
{\csc Abstract.}
This paper develops a nonparametric density estimator
with parametric overtones. 
Suppose $f(x,\theta)$ is some family of densities,
indexed by a vector of parameters $\theta$. 
We define a local kernel smoothed likelihood function 
which for each $x$ can be used to estimate the 
best local parametric approximant to the true density. 
This leads to a new density estimator
of the form $f(x,\hatt\theta(x))$, thus inserting the best
local parameter estimate for each new value of $x$. 
When the bandwidth used is large this amounts to 
ordinary full likelihood parametric density estimation, 
while for moderate and small bandwidths the method
is essentially nonparametric, using only 
local properties of data and the model.
Alternative ways more general than via 
the local likelihood are also described.   
The methods can be seen as ways of nonparametrically smoothing
the parameter within a parametric class. 

Properties of this new semiparametric estimator are investigated.
Our preferred version has approximately the same variance as the ordinary
kernel method but potentially a smaller bias. 
The new method is seen to perform better than the traditional kernel method
in a broad nonparametric vicinity of the parametric model employed,
while at the same time being capable of not losing much in precision
to full likelihood methods when the model is correct. 
Other versions of the method are equivalent to using 
particular higher order kernels in a semiparametric framework.
The methodology we develop can be seen 
as the density estimation parallel to local likelihood
and local weighted least squares theory in nonparametric regression. 

\smallskip\noindent
{\csc Key words:} 
{\sl
bias reduction, 
density estimation, 
kernel smoothing, 
local likelihood, 
local modelling,
parameter smoothing,  
semiparametric estimation}
\smallskip}}

\bigskip
{\bf 1. Introduction and summary.} 
Let $X_1,\ldots,X_n$ be independent and identically
distributed with density $f$. The traditional
kernel estimator of $f$ is $\tilda f(x)=n^{-1}\sumin K_h(x_i-x)$,
where $K_h(z)=h^{-1}K(h^{-1}z)$ and $K(.)$ is some chosen unimodal
density, symmetric about zero. The basic properties of $\tilda f$ are
well known, and under smoothness assumptions these include 
$$\eqalign{
\E\tilda f(x)&=f(x)+\half\sigma_K^2h^2f''(x)+O(h^4) \cr
{\rm and} \quad
\Var\,\tilda f(x)&=R(K)(nh)^{-1}f(x)-n^{-1}f(x)^2+O(h/n), \cr} \eqno(1.1)$$
where $\sigma_K^2=\int z^2K(z)\,\d z$ and $R(K)=\int K(z)^2\,\d z$.
See Scott (1992, Chapter 6) or Wand and Jones (1995, Chapter 2), for example. 

Our aim in this paper is to propose and investigate 
a class of semiparametric competitors which have precision 
comparable to that of $\tilda f$ but sometimes better. 
For any given parametric family 
$f(\cdot,\theta)=f(\cdot,\theta_1,\ldots,\theta_p)$
and for each given $x$ we will present ways of estimating 
the locally best approximant to $f$, and then use 
$$\hatt f(x)=f(x,\hatt\theta_1(x),\ldots,\hatt\theta_p(x)). \eqno(1.2)$$ 
Thus the estimated density at $x$ employs a parameter 
value which depends on $x$
and whose choice is to be tailored to good estimation at $x$.
In other words the method amounts to a version of 
nonparametric parameter smoothing within the given parametric class. 

\subsection
{\csc 1.1. Local likelihood for densities.} 
A central idea in our paper is the construction of a 
{\it local likelihood function} for density estimation. 
Local likelihood ideas have been employed in non- and semiparametric
regression for some time, see Section 1.2, but the concept is
far less immediate in the present context of density estimation.   
Around each given $x$ we define the local log-likelihood to be  
$$\eqalign{
L_n(x,\theta)&=\int K_h(t-x)\{\log f(t,\theta)\,\d F_n(t)
        -f(t,\theta)\,\d t\} \cr
&=n^{-1}\sumin K_h(x_i-x)\log f(x_i,\theta)
        -\int K_h(t-x)f(t,\theta)\,\d t, \cr}\eqno(1.3)$$
writing $F_n$ for the empirical distribution function. 
When $h$ is large this is close to the constant $K(0)h^{-1}$ times 
the ordinary, normalised log-likelihood function 
$n^{-1}\sumin\log f(x_i,\theta)-1$, 
and maximising the (1.3) function 
with respect to the parameters becomes equivalent to ordinary full 
maximum likelihood estimation. When $h$ is moderate or small,
however, maximising $L_n(x,\theta)$ will be seen to be a 
fruitful way of obtaining an estimate of the best local approximant 
to $f$. This is made clear in Section 2.  

A related and in fact more general apparatus is as follows. 
Decide on suitable weight functions $v_j(x,t,\theta)$,  
$j=1,\ldots,p$, guidelines for which will be discussed later, 
and let $\hatt\theta(x)$ be defined as the solution to
the $p$ equations   
$$\eqalign{
V_n(x,\theta)&=\int K_h(t-x)v(x,t,\theta)\{\d F_n(t)-f(t,\theta)\,\d t\} \cr
&=n^{-1}\sumin K_h(x_i-x)v(x,x_i,\theta)
        -\int K_h(t-x)v(x,t,\theta)f(t,\theta)\,\d t=0. \cr}\eqno(1.4)$$ 
Maximising the (1.3) function amounts to solving (1.4) 
with $v(x,t,\theta)=u(t,\theta)={\dell\over \dell\theta}\log f(t,\theta)$,  
the $p\times1$ score function of the model, with one component 
$u_j(x,\theta)$ per parameter. 
The generalisation is analogous to that of M-estimation 
over maximum likelihood estimation in ordinary estimation theory. 

This strategy, with (1.4) or its special case (1.3), 
gives $\hatt\theta(x)$ and in the end (1.2). 
We call this {\it local parametric estimation} of the 
density $f$, hence the title of our paper. 
An attractive motivation for this approach is that as 
$h\arr\infty$, $\hatt f$ tends to a global 
parametric fit of the model $f(\cdot,\theta)$. 
As in other attempts at semiparametric density estimation,
cf.~references mentioned below, 
our methodology should be particularly useful when $f$ exhibits 
small or moderate departures from a standard parametric form. 
But $f(\cdot,\theta)$ need not even be a crude model for the data because, 
if not, $h$ will be chosen small, and local properties of $\hatt f$ 
will largely be divorced from global properties of $f(\cdot,\theta)$.
Thus we view our method as a `continuous bridge' between
fully parametric and fully nonparametric options. 

The local likelihood function is more fully 
motivated --- in several ways --- in Section 2,
and a connection is also established to the dynamic likelihood methods 
for nonparametric hazard rate estimation of Hjort (1991, 1996a). 
Apart from the local likelihood connection, 
we note that the (1.4) type approach is natural 
in that a weighted difference of $\d F_n(t)-f(t,\theta)\,\d t$, 
which in the limit is a weighted difference of 
$\{f(t)-f(t,\theta)\}\,\d t$, is set to zero.

The new estimator can and will be motivated also 
on the grounds of performance, of course. 
We start our investigation of the large sample properties of $\hatt f(x)$
in Section 3, with concentration on one-parameter local fits. 
This is extended in Section 4 to the multiparameter case, 
with particular focus on two parameters. 
The two parameter case affords an attractive 
simplification of $O(h^2)$ bias, 
and forms our favoured class of locally parametric density estimators.
It turns out that the bias and variance properties of $\hatt f$ are 
remarkably comparable to those of the classic estimator $\tilda f$. 
For many situations it will be seen that 
$$\eqalign{
\E\hatt f(x)&=f(x)+\half\sigma_K^2h^2b(x)+O(h^4+n^{-1}) \cr
{\rm and} \quad
\Var\,\hatt f(x)&=R(K)(nh)^{-1}f(x)-n^{-1}f(x)^2+O(h/n), \cr}\eqno(1.5)$$
just as in (1.1), but with a bias factor function $b(x)$ 
related to but different from $f''(x)$,  
with characteristics inherited from the parametric class 
and the weight functions used. 
To the order of approximation used the variance is simply the same,
regardless of parametric family and of $v(x,t,\theta)$. 
The statistical advantage will be that for many $f$'s,
typically those lying in a broad nonparametric neighbourhood 
of the parametric $f(\cdot,\theta)$,
$b(x)$ will be smaller in size than $f''(x)$ for most $x$. 
It should also be the case
that in aiming for improved performance by choice of $f(\cdot,\theta)$ 
we will rarely lose too much in performance terms in the sense that 
$b(x)$ should not be too much greater than $f''(x)$ on occasions when 
$f(\cdot,\theta)$ is a totally inappropriate global model.

In Section 4 it is also shown that a bias of the potentially 
smaller size $O(h^4)$ is attainable if the vehicle model 
has three or four parameters and the underlying true density
is sufficiently smooth. 
This is achieved without having to 
(explicitly) resort to higher order kernels. 
Our method is, however, in its kernel-dependent quantities, 
asymptotically equivalent to a particular class
of higher order kernels which are of the form 
a suitable polynomial times $K$. The same higher order kernels arise
in local polynomial regression (see Section 1.2 below),  
but we stress that this result is consequent on the number
of parameters fitted and not on using any particular form of local
parameterisation (which shows up only in the bias factor).  
We conjecture that the same is true in the 
local least squares regression context.   
Thus locally smoothing a three- or four-parameter model 
leads to a superior asymptotic performance.
We nevertheless favour two-parameter families for 
their comparative simplicity conceptually and computationally, 
and with experience of higher order kernels
raising doubts about the transfer of such 
asymptotic advantages to finite sample practice (Marron and Wand, 1992).

A variety of particular examples is discussed in Section 5.
These are not practical examples but rather features and properties 
of interesting special cases of our methodology. 
Particular attention is given to the case of an estimated  
`running normal' density and to estimates that incorporate
local modelling of level, slope and curvature. 
Sections 6 and 7 provide further extensions of the earlier theory. 
In Section 6, we present results on the boundary behaviour of our
estimators, and note attractive properties thereof. 
Section 7 indicates extensions to the multivariate case,
where the new method could prove to be particularly useful, 
since the ordinary methods are problematic in higher dimensions. 
In Section 8 we discuss some other issues such as 
automatic bandwidth selection and inspection of `running parameters',
while our conclusions are offered in Section 9.
Our focus throughout this paper 
is on intuitive and theoretical considerations. Implementation
issues and comparative work are left to future studies.

\subsection
{\csc 1.2. Related work.}                               
In nonparametric regression, there has been much recent interest 
in fitting polynomial functions locally; relevant references include 
Fan (1992, 1993), Fan and Gijbels (1992, 1996),
Hastie and Loader (1996), Ruppert and Wand (1994),
building on earlier work of Stone (1977) and Cleveland (1979). 
This has been done by local least squares, which is a normal 
error distribution version of local likelihood fitting,
see Tibshirani and Hastie (1987), Staniswalis (1989), 
Jones and Hjort (1994), and Fan, Heckman and Wand (1995). 
Local linear fitting is particularly attractive.
It affords asymptotic bias depending only on the second derivative 
of the regression function, 
without sacrificing anything in terms of variance (this is not at all
trivial to achieve, cf.~Jones, Davies and Park, 1994). 
It also automatically has very good boundary properties. 
Higher degree polynomials behave rather like
`higher order' kernels. In the large bandwidth limit, the parametric
form approached is, of course, a global polynomial regression. 
Given the large impact of these methods in regression, 
it is natural to ask if parallel methods can be invented 
for density estimation. It is indeed an aim of this paper to
provide such a methodology. 

At around the same time as we were developing our ideas, 
Loader (1996) independently proposed a version of 
local likelihood density estimation. 
A key component is specification of an appropriate likelihood function, 
and Loader's definition is indeed similar to our (1.3).
Loader uses his definition to fit local polynomials to the log density, 
perhaps the most immediate analogue of the regression work.
Our motivation differs from Loader's in preferring to work 
with more general local parametric models, seeking semiparametric
density estimators, with standard parametric models as limiting cases.  
However, our methodology covers interesting 
non-standard parametric forms, 
and other local estimation methods, as well. 
We arrived at (1.3) and its relative (1.4) 
partly via the hazard rate case, 
for which local likelihood specification 
is more immediate, see Hjort (1991, 1996a), and partly 
via local weighting of the $\d F_n(t)-f(t,\theta)\,\d t$ difference;
see Section~2. 

Some semiparametric density estimators already exist.
Our approach has similar intentions to that of Copas (1995), 
but ours appears to be both simpler and more general. 
A semiparametric method which works by multiplying an
initial parametric description with a nonparametric kernel-type
estimate of the necessary correction factor is developed in
Hjort and Glad (1995). Their estimator also has properties (1.5),
but with yet another $b(x)$ bias factor function. 
Another similarly spirited method 
consists in using an estimated orthogonal expansion for this
multiplicative correction factor; 
see, for example, Hjort (1986, Chapter 5), Buckland (1992) 
and Fenstad and Hjort (1996).  
An initial nonparametric estimator ``corrected''
towards the parametric is the topic of recent work 
of Efron and Tibshirani (1996). 
These authors also note the role of backfitting as in Hastie
and Tibshirani (1990) in a similar context. 
Various semiparametric density estimators 
of Bayesian flavour are discussed in Hjort (1995).
Earlier work, somewhat less attractively
involving an extra parameter in a linear combination 
of parametric and nonparametric estimators, 
includes Schuster and Yakowitz (1985) and Olkin and Spiegelman (1987). 
Jones (1993a) argues that (the natural variance-corrected version of) 
the kernel density estimator can itself 
be thought of as a semiparametric estimator.

\bigskip
{\bf 2. Local likelihood for density estimation.}
This section gives support for 
the local parametric estimation method of (1.2)--(1.3). It first relates
the method to a well-defined local statistical Kullback--Leibler type 
distance function from the true density to the parametric approximant.
This is followed by a connection
to similar concepts for hazard rate estimation in survival data.
Finally included in this section are alternative motivations, 
also of others, for considering the same definition of local likelihood. 

\subsection
{\csc 2.1. Local parametric approximation.}                             
To explain why maximising (1.3) is a good idea, note first that 
$$L_n(x,\theta)\arr_p\lambda(x,\theta)
=\int K_h(t-x)\{f(t)\log f(t,\theta)-f(t,\theta)\}\,\d t $$
as $n$ grows. 
The maximiser $\hatt\theta(x)$ hence aims at the parameter value $\theta_0(x)$ 
that maximises $\lambda(x,\theta)$. This is a well-defined statistical
quantity in that it minimises the distance  
$$d[f,f(\cdot,\theta)]=\int K_h(t-x)[f(t)\log{f(t)\over f(t,\theta)}
        -\{f(t)-f(t,\theta)\}]\,\d t \eqno(2.1)$$
between true density (which need not belong to the parametric class
under consideration) and approximating parametric density. 
Noting that the Kullback--Leibler distance from 
$f$ to $f_\theta$ can be written
$$\int f(t)\log\{f(t)/f(t,\theta)\}\,\d t
  =\int\bigl[f(t)\log{f(t)\over f(t,\theta)}
        -\{f(t)-f(t,\theta)\}\bigr]\,\d t, $$
we see that (2.1) is a version of the same, locally weighted around $x$.
These arguments show that using (1.2) with (1.3), 
which is (1.4) with weight function chosen to be 
the score function $u(t,\theta)$, 
aims at the best local parametric approximant to the true $f$. 
Note also that if $f$ is not far from $f(\cdot,\theta)$, then 
$d[f,(f\cdot,\theta)]\simeq
        \half\int K_h(t-x)\{f(t)-f(t,\theta)\}^2/f(t)\,\d t$. 
An alternative $L_2$-based local distance measure is briefly
discussed in Section 5.6. 

\subsection
{\csc 2.2. The hazard connection.}
For a moment, consider survival data on $[0,\infty)$, and
switch attention from density $f(t,\theta)$ and cumulative 
distribution $F(t,\theta)$ to survival function 
$S(t,\theta)=1-F(t,\theta)$ and, particularly, hazard function 
$\alpha(t,\theta)=f(t,\theta)/S(t,\theta)$.
The likelihood is $\prod_{i=1}^n\alpha(t_i,\theta)\allowbreak S(t_i,\theta)$ 
so that the log-likelihood, 
after a little manipulation, and disregarding a multiplier
of $n$, takes the form
$\int\{\log\alpha(t,\theta)\allowbreak\,\d F_n(t)
        -S_n(t)\alpha(t,\theta)\,\d t\}$, 
where $S_n(t) = 1-F_n(t)$ is the proportion of individuals 
still at risk just prior to time $t$.
The kernel smoothed local log-likelihood for the model 
at location $x$ is, therefore, 
$$L_{0,n}(x,\theta)=\int K_h(t-x)\{\log\alpha(t,\theta)\,\d F_n(t)
        -S_n(t)\alpha(t,\theta)\,\d t\}. \eqno(2.2)$$
This local likelihood for hazard models is 
well-motivated and explored in Hjort (1991, 1996a). 
Note that 
$$L_{0,n}(x,\theta)\arr_p\lambda_0(x,\theta)
=\int K_h(t-x)\Bigl\{f(t)\log{f(t,\theta)\over S(t,\theta)}
        -S(t){f(t,\theta)\over S(t,\theta)}\Bigr\}\,\d t. $$
Maximising $L_{0,n}(x,\theta)$ aims at the best local approximant 
in the sense of minimising the local distance function
$$\eqalign{
  d_0[f,f(\cdot,\theta)]
  &=\int K_h(t-x)\Bigl[f(t)\Bigl\{\log{f(t)\over S(t)}
        -\log{f(t,\theta)\over S(t,\theta)}\Bigr\} \cr
  &\qquad -S(t)\Bigl\{{f(t)\over S(t)}-{f(t,\theta)\over S(t,\theta)}\Bigr\}
        \Bigr]\,\d t. }$$
This underlies the theory of locally parametric nonparametric
hazard rate estimation, and is as in Hjort (1996a, Sections 2 and 3), 
but now suitably re-expressed as a distance between densities and 
not hazards. 

To see a connection from this context to density estimation, 
put in $\alpha(t,\theta)=f(t,\theta)/S(t,\theta)$ to see 
$$\eqalign{
  L_{0,n}(x,\theta)
  &=\int K_h(t-x)[\{\log f(t,\theta)
  -\log S(t,\theta)\}\,\d F_n(t) \cr
  &\qquad -S_n(t)f(t,\theta)/S(t,\theta)\,\d t]. }$$
Now replace $S(t,\theta)$ here with the estimate $ S_n(t)$ 
(this step will be discussed in Section 2.3). 
This leads to 
$$\int K_h(t-x)[\{\log f(t,\theta)-\log S_n(t)\}\,\d F_n(t)
        -f(t,\theta)\,\d t], $$
and since the $\log S_n(t)$ term is immaterial this is the same as 
$L_n(x,\theta)$ of (1.3). 
We point out that the hazard connection makes it 
clear how censoring can be coped with also, 
see Hjort (1996a). 

\subsection
{\csc 2.3. Justification of (1.3) as local log-likelihood.}
We think of (1.3) as the {\it local log-likelihood}, 
or local kernel smoothed log-likelihood, 
for the model at $x$. The main justification for this 
is via the best local approximation framework laid out in Section 2.1
above, combined with the appealing feature that large bandwidths
lead back to global likelihood analysis, and not least 
with the fact that the method works, as this paper demonstrates.  
We also know of four additional justifications for the (1.3)
construction. 

The first completes the argument of Section 2.2.  
One can argue that the insertion of $S_n(t)$ for $S(t,\theta)$ 
here should not alter things very much since $S_n(t)$ 
is a more precise estimate 
than is any local parameter estimate (or hence local density estimate) 
for its population version. 
Indeed, $S_n(t)$ has mean squared error of order $n^{-1}$, which is
insignificant compared with the mean squared error of our density estimate
which, it will turn out, will be $O(n^{-4/5})$.

But what of a more direct local likelihood argument? 
The na\"\i ve local log-likelihood $\int K_h(t-x)\log f(t,\theta)\,\d F_n(t)$
does not work, as inspection in the normal case
pedagogically reveals, for example. Similarly the na\"\i ve nonparametric
log-likelihood $\int\log f(t)\,\d F_n(t)$ 
has problems, whether kernel smoothed or not; it can be made 
infinite by putting infinite spikes at the data points. 
Loader (1996) argues that the log-likelihood is truly
$\int\log f(t)\,\d F_n(t)-\int f(t)\,\d t$
(think of likelihood estimation of a Poisson intensity function)
but the final term is normally discarded since it takes the value one. 
Leaving the second term in and then localising by kernels 
yields precisely (1.3) again. 

Another argument stems from J.B.~Copas (personal communication).
Note first that the derivative of the simplistic 
$\int K_h(t-x)\log f(t,\theta)\,\d F_n(t)$ is 
$\int K_h(t-x)u(t,\theta)\,\d F_n(t)$,  
which does not have expectation zero, even under model conditions.  
To remedy this, subtract its expectation, which is 
$\int K_h(t-x)u(t,\theta)f(t)\allowbreak\,\d t$. 
Or, at least, if we approximate this last $f(t)$
by $f(t,\theta)$ we obtain the score function case of 
$V_n(x,\theta)$ of (1.4), and hence motivate $L_n(x,\theta)$
at (1.3) once more. 
(Copas's, 1995, suggestion differs from this. 
The current version replaces Copas's expression (7),
$w(x)\log f(x,\theta)+\{1-w(x)\}\log B(\theta)$ in Copas's notation, 
by $w(x)\log f(x,\theta)+B(\theta)-1$.)

Comments from a referee triggered the following fourth 
justification of (1.3). This is interesting in that it 
connects the density estimation problem 
to the more well-developed local likelihood methodology 
for nonparametric regression. It is based on a discretisation argument: 
split the data region into small intervals $D_1,\ldots,D_m$ of lengths
$d_1,\ldots,d_m$,
and let $s_1,\ldots,s_m$ be the number of points falling in each.
Modelling the $s_j$s as independent Poisson variables with parameters 
$\gamma\pi_j(\theta)$, where 
$\pi_j(\theta)=\int_{D_j}f(x,\theta)\allowbreak\,\d x$,
gives (omitting an additive constant) the log-likelihood 
$\sum_{j=1}^m\{-\gamma\pi_j(\theta)+s_j\log\gamma+s_j\log\pi_j(\theta)\}$. 
Conditioning the Poisson model on $\sum_{j=1}^m s_j\allowbreak=n$,
which is also the maximum likelihood estimate of $\gamma$, 
corresponds to the multinomial model for the $s_j$ counts. 
This formal equivalence to the Poisson model 
was exploited in Lindsey (1974) and more recently in 
Efron and Tibshirani (1996). 

The present point is that there is a well-established way of 
localising such a likelihood, 
see Tibshirani and Hastie (1987), 
Jones and Hjort (1994), 
Fan, Heckman \& Wand (1995)
and Fan and Gijbels (1996), 
since it has been made to belong to nonparametric smoothing 
of Poisson parameters rather than density estimation. This gives 
$$\tilda L_n(x,\theta)
        =\sum_{j=1}^mK_h(x-x_{(j)})[-\gamma\pi_j(\theta)
                +s_j\log\{\gamma\pi_j(\theta)\}] $$
where $x_{(j)}$ is a convenient point in $D_j$. Taking a fine limit,
via $\pi_j(\theta)\simeq d_j f(x_{(j)},\theta)$, leads to 
$$ \tilda L_n(x,\theta)\simeq -\gamma\int K_h(x-t)f(t,\theta)\,\d t
        +(\log\gamma)\,n\tilda f(x)+\sumin K_h(x-x_i)\log f(x_i,\theta) .$$
Putting $\gamma=n$ here,
as suggested by the original Poisson connection, gives (1.3) again. 
The connection isn't quite as clear-cut, however, 
since the maximiser is $\hatt\gamma_\theta=n\tilda f(x)/(K_h*f_\theta)(x)$
which still depends on $\theta$, and this
delivers another proposal, namely the profile log-likelihood 
$$L_n^*(x,\theta)=-n\tilda f(x)\,\log\Bigl\{\int K_h(x-t)
        f(t,\theta)\,\d t\Bigr\}+\sumin K_h(x-x_i)\log f(x_i,\theta). $$
We would still have $\hatt\gamma_\theta=n(1+O(h^2))$ for small $h$ and
for the $\theta$s
of interest, however, leading again to (1.3). 

See Jones (1995) for more on discretised forms of local likelihood.
\eject 

\bigskip
{\bf 3. Large sample properties.}

\subsection
{\csc 3.1. $h$ fixed, large $n$.}
Let $\theta$ be $p$-dimensional in this subsection. 
Estimating $\theta$ by solving (1.4) is like M-estimation,
with the extra complication that we do not assume the true $f$ to
belong to the parametric $f(\cdot,\theta)$ class. 
For simplicity suppress the fixed $x$ and write $v(t,\theta)=v(x,t,\theta)$
for the $p$ weight functions. 
Assume that
$$V(x,\theta)=\int K_h(t-x)v(t,\theta)\{f(t)-f(t,\theta)\}\,\d t=0 \eqno(3.1)$$
has a unique solution $\theta_0=\theta_0(x)$
(which also depends on $h$, held fixed here).  
This essentially says that $f(t)$ should be 
within reach of $f(t,\theta)$ as $\theta$ varies and that the 
$p$ functions $v_j(t,\theta)$ should be functionally independent;  
see the examples of Section 5.
That $V_n(x,\theta_0)$ has mean zero plays a role in 
developing the following facts.  
Firstly, $\hatt\theta(x)$ converges to this best local parameter 
$\theta_0(x)$ in probability. 
In the score function case $v=u$ this is also the parameter minimising (2.1). 
Secondly, 
$$(nh)^{1/2}\{\hatt\theta(x)-\theta_0\}\arr_d
        \normal_p\{0,J_h^{-1}M_h(J_h')^{-1}\}, \eqno(3.2)$$
where 
$$\eqalign{
J_h&=\int K_h(t-x)\bigl[v(t,\theta_0)u(t,\theta_0)'f(t,\theta_0)
        +v^*(t,\theta_0)\{f(t,\theta_0)-f(t)\}\bigr]\,\d t, \cr
M_h&={\rm VAR}_f\{h^{1/2}K_h(X_i-x)v(X_i,\theta_0)\} \cr 
   &=\int hK_h(t-x)^2v(t,\theta_0)v(t,\theta_0)'f(t)\,\d t 
        - h\xi_h\xi_h', \cr}$$
and $\xi_h=\int K_h(t-x)v(t,\theta_0)f(t)\,\d t$. 
Again $u(t,\theta)$ is the model's score function while 
$v^*(t,\theta)$ is the $p\times p$ matrix of derivatives of the 
$v_j(t,\theta)$ functions.
Proving these claims is not very difficult,
using variations of arguments used to prove asymptotic normality  
of M-estimators; see Section 8.4 for relevant details 
and an additional result. By the delta method
$$(nh)^{1/2}\{\hatt f(x)-f(x,\theta_0)\}
\arr_d\normal\{0,f(x,\theta_0)^2
        u(x,\theta_0)'J_h^{-1}M_h(J_h')^{-1}u(x,\theta_0)\}. \eqno(3.3)$$

\subsection
{\csc 3.2. Decreasing $h$.} 
The (3.3) result is valid for a fixed positive $h$. 
We are also interested in being increasingly fine-tuned about
$h$ as $n$ grows. Observe that, as $h\arr0$,  
$$\int K_h(t-x)g(t)\,\d t=g(x)+\half\sigma_K^2h^2g''(x)+O(h^4) \eqno(3.4) $$
for each smooth $g$ function, by a standard simple Taylor series argument. 
Using this in conjunction with (3.1) 
shows that $f(x,\theta_0(x))-f(x)=O(h^2)$ in general. Indeed, 
$$v_{j,0}(x)\{f_0(x)-f(x)\}=\half\sigma_K^2h^2
        \{v_{j,0}(f-f_0)\}''(x)+O(h^4) \eqno(3.5)$$
under smoothness assumptions on $f$ and the weight functions,
writing $f_0(x)=f(x,\theta_0)$, $v_{j,0}(x)=v_j(x,\theta_0)$ and so on
(and where $\theta_0=\theta_0(x)$ also depends on $x$). 
Furthermore $(v_{j,0}f_0)''(x)$, for example, means the second $x$-derivative 
of the $v_j(x,\theta)f(x,\theta)$ function, and then inserted 
the parameter value $\theta=\theta_0(x)$. 
Under mild regularity assumptions this also implies 
$$\E\hatt f(x)=f(x)+\half\sigma_K^2h^2b(x)+O(h^4+(nh)^{-1}), $$
where the precise nature of the $b(x)$ function will be quite
important and will be analysed more later. 

We need to assess the size of $J_h^{-1}M_h(J_h')^{-1}$ of (3.2),
and of the variance appearing in (3.3), when $h$ tends to zero.
To this end it proves to be convenient to reparametrise quantities 
in $J_h$ and $M_h$. Rewrite $f(t,\theta_0)$ as $f(t-x,\psi_0)$,
where the new parameters $\psi$ are easily related to the
old parameters $\theta$ and we note that the first element of 
$\hatt\psi$ is the only one directly specifying $\hatt f(x)$. 
Also, replace $u(t,\theta_0)$ and $v(t,\theta_0)$ by
$u_h(h^{-1}(t-x),\psi_0)$ and 
$v_h(h^{-1}(t-x),\psi_0)$ respectively, 
the subscript $h$ referring to dependence of $u_h$ and $v_h$ on
$h$ to accommodate the $h^{-1}$ attached to $t-x$. 
(For an example, reparametrise $\theta_1+\theta_2t+\theta_3t^2$ 
to $\psi_1+\psi_2hz+\psi_3h^2z^2$ where $z=(t-x)/h$.)
We then find that 
$$\eqalign{
J_h&=f_0(x)\int K(z)v_h(z,\psi_0)u_h(z,\psi_0)'\,\d z+O(h^2), \cr
M_h&=f(x)\int K(z)^2v_h(z,\psi_0)v_h(z,\psi_0)'\,\d z
        -h\xi_0\xi_0'f(x)^2+O(h^2), \cr} \eqno(3.6)$$
where $\xi_0=\int K(z)v_h(z,\psi_0)\,\d z$. 

\subsection
{\csc 3.3. The one-parameter case.}
Let $f(x,\theta)$ have just one parameter and let the weight function 
$v(t,\theta)$ be smooth and non-zero at $x$. From 
(3.5) and previous arguments one finds 
$$b(x)=f''(x)-f_0''(x)+2\{v_0'(x)/v_0(x)\}\{f'(x)-f_0'(x)\}, \eqno(3.7)$$
differing from the kernel estimator's bias factor 
$f''(x)$ by a term depending on properties of $f(\cdot,\theta)$. 
If $f_0=f$, that is, if we are working with the 
correct parametric class, then $b(x)=0$. Otherwise, 
(3.7) should be small when $f_0$ is close to $f$, 
and perhaps not too large in absolute value even when $f$
and $f_0$ differ considerably. 
Notice that the expression for $b(x)$ 
simplifies when the weight function used is $v(t,\theta)=1$.
It also simplifies in the multi-parameter case of the next section.
An expression for the variance is found from (3.3) and (3.6). 
Assuming that $v_h(z)$ and $u_h(z)$ are of the form $c+O(hz)$ for small $h$, 
the weight function as well as other traces of the parametric model
are seen to cancel out, for the leading terms, and the result is 
$$\Var\,\hatt f(x)=R(K)(nh)^{-1}f(x)-n^{-1}f(x)^2+O(h/n). \eqno(3.8)$$
That is, the variance is the same, to the order of approximation used, 
as that of the ordinary kernel density estimator. 

\bigskip
{\bf 4. The multiparameter case.}
In this section, let the parametric model be 
$f_{\theta}(x)=f(x,\theta_1,\ldots,\theta_p)$ with $p\ge2$.
The results we shall obtain for approximate biases and 
variances again hold under suitable regularity assumptions,
including permission to interchange limits and expectation.
That these are met can be checked directly for the most
important special cases, like those listed in Section 5. 

\subsection
{\csc 4.1. The bias.}
We have $\E\hatt f(x)=f(x,\theta_0)+O((nh)^{-1})$ again,
and the $p$ equations 
$\int K_h(t-x)v_j(t,\theta_0)\{f(t)-f(t,\theta_0)\}\,\d t=0$
can be used to see how far 
$f(x,\theta_0)$ is from $f(x)$. From (3.5) it is seen that 
$$f(x,\theta_0)-f(x)=\half\sigma_K^2h^2[f''(x)-f_0''(x)
        +2\{v_{j,0}'(x)/v_{j,0}(x)\}\{f'(x)-f_0'(x)\}]+O(h^4) $$
for each $j$, under smoothness assumptions. 
Since there are $p\ge2$ equations giving the 
$h^2$ coefficient this can only hold when $f'(x)-f_0'(x)=o(1)$
as $h\arr0$. This is not in general true in the one-parameter case, 
and is the cause of the extra term making up (3.7). 
For $p\ge2$, however, we have
$$\E\hatt f(x)=f(x)+\half\sigma_K^2h^2\{f''(x)-f_0''(x)\}
        +O(h^3+(nh)^{-1}). \eqno(4.1)$$
Introduction of further local parameters has simplified the bias to
depending solely on $f''(x)-f_0''(x)$. This is appealingly interpretable.
The bias is of a familiar second derivative, local curvature, type, 
and the way in which closeness of $f_0$ to $f$ affects the 
bias is abundantly clear.

But the above remarks are really most relevant to the case of two parameters
exactly. For $p\ge3$, an extension of the above argument shows that $(f-f_0)''$
is also $o(1)$. To see this, write 
$g_r \ident (f-f_0)^{(r)} \simeq \sum_{i=0}^{4-r}a(r,i) h^i$
for $r=0,\ldots,4$. Then look at the general equations 
governing asymptotic bias and equate terms in powers of $h$. 
These are
$$g_0 + \half k_2h^2(v_{j,0} g_0)''/v_{j,0} 
+\frac1{24}k_4h^4(v_{j,0} g_0)^{(4)}/v_{j,0}
+\frac1{720}k_6h^6(v_{j,0} g_0)^{(6)}/v_{j,0}=0, $$
for $j=1,\ldots,p$, where we write $k_j=\int z^jK(z)\,\d z$;
in particular $k_2=\sigma_K^2$.  
For instance, when $p=3$, a little manipulation yields
$a(0,i) = 0 = a(j,k)$ for $i=0,1,2,3$, $j=1,2$, $k=1,2$. 
Also of importance are
$a(0,4)+{1\over 2}k_2a(2,2)+{1\over 24}k_4a(4,0)=0$ and
$k_2a(1,2)+{1\over 6}k_4a(3,0)=0$. 
To make further progress, we need to consider the $h^6$ term which involves
three equations in four unknowns, 
where in particular two of these, say $t$ and $u$ satisfy
$$t = \frac12 k_2 a(0,4) +\frac14 k_4a(2,2) 
        + \frac1{48}k_6a(4,0) 
        ~{\rm and}~ 
        u=\frac16 k_4a(1,2) + \frac1{36}k_6a(3,0).$$
We can thus write $t=Au$ for appropriate $A$ and hence find that
$$a(0,4) = {k_2k_6-k_4^2\over k_4 - k_2^2}
        \{\frac1{24} a(4,0) - \frac1{18} A a(3,0)\}.$$
Reinterpreting this in bias terms results in 
$$\E\hatt f(x)=f(x)- 
{k_2k_6-k_4^2 \over k_4-k_2^2}h^4 
\bigl[\frac1{24} \{f^{(4)}(x)-f_0^{(4)}(x)\} -
\frac1{18} A \{f^{(3)}(x)-f_0^{(3)}(x)\} \bigr]+o(h^4), \eqno(4.2)$$
where, being explicit about $A$,
$A$ is the solution to the system of equations
$x v_{j,0}+y v'_{j,0}+Av''_{j,0}=-v'''_{j,0}$ 
for $j=1,2,3$. 

Increasing $p$ from 3 to 4 results in the considerable simplification that
the term involving $A$ in (4.2) disappears due to being able to set $t=u=0$
so that we then have
$$\E\hatt f(x)=f(x)-
        \frac1{24}h^4{k_2k_6- k_4^2 \over k_4 -k_2^2} 
        \{f^{(4)}(x)-f_0^{(4)}(x)\}+o(h^4).  \eqno(4.3)$$ 
Therefore, one gets an exact parallel of properties of the
local polynomial regression referred to in Section 1.2;  
see Ruppert and Wand (1994) and Fan and Gijbels (1996). 
Fitting one or two parameters, using a second-order kernel $K$, 
corresponds to $O(h^2)$ bias,
with two parameters exhibiting advantages in terms of simplicity, 
and the same goes for local constant and linear regression. 
Three and four local parameters yield $O(h^4)$ bias, 
as do local quadratic and cubic regressions, 
and four parameters affords a simple dependence on
$(f-f_0)^{(4)}$. And, we conjecture, so on 
(sufficient smoothness of $f$ and the parametric densities used permitting). 
An important point emerging here is that we have not had to impose any
particular local parametric form to achieve this behaviour. 
Rather it is a consequence of the number of local parameters fitted. 
See also Sections 4.2--4.3 below.
Since the practical value of these asymptotic results is perhaps
dubious, we prefer to concentrate on the two parameter case and consequent
improvements in leading constant rather than rate, allied with more obvious
practical interpretation.

\subsection
{\csc 4.2. The variance.}
We use (3.3) with (3.6), assuming, as is reasonable, 
that the $v_h(z,\psi_0)$ and $u_h(z,\psi_0)$ functions 
are of the form $c_1+c_2(hz)+c_3(hz)^2+\cdots$ for small $h$, 
and that there is at least one nonzero $c_i$ coefficient 
in each of the vectors $v_h$ and $u_h$. 
It should be no surprise that $v$ and $u$ functions can be subjected to 
arbitrary linear transformations without effect on the 
resulting estimates, and it is easy to see by consideration
of $u'J_h^{-1}M_h(J_h')^{-1}u$ and (3.6) 
that the variance is unaffected by this. 
As far as this asymptotic assessment is concerned, therefore,
where $h\arr0$ and $nh\arr\infty$,  
it follows that we can replace both 
$v$ and $u$ by the canonical function 
$$V_h(z)=(1,hz,h^2z^2,\ldots,h^{p-1}z^{p-1}). $$
Thence, from (3.3) and (3.6), we see that 
$$\Var\,\hatt f(x)=(nh)^{-1}f(x)\tau(K)^2-f(x)^2/n+O(h+h/n), \eqno(4.4)$$
where, letting $e_1=(1,0,\ldots,0)'$,   
$$\tau(K)^2=e_1'
        \Bigl(\int KV_hV_h'\,\d z\Bigr)^{-1}
        \Bigl(\int K^2V_hV_h'\,\d z\Bigr)
        \Bigl(\int KV_hV_h'\,\d z\Bigr)^{-1}e_1. $$

A particularly natural local parameterisation takes 
$f(t,\theta)$ as $\exp(\sum_{j=0}^{p-1}\theta_jt^j)$
so that $u(t,\theta)=(1,t,\ldots,t^{p-1})$. 
This is the special case --- with $v=u$ --- explored 
by Loader (1996), who gives essentially the same 
variance expression as above. But we must emphasise 
that this variance result also holds for {\it any} (sensible) local 
parametrisation and not just for Loader's: it is purely 
a consequence, as is the kernel-dependent part of the bias,
of the {\it number} of local parameters fitted. 

\subsection
{\csc 4.3. Two, three, four parameters.}
In the case of two parameters, (4.4) simply reduces to 
$$\Var\,\hatt f(x)=(nh)^{-1}f(x)R(K)-n^{-1}f(x)^2+O(h/n). \eqno(4.5)$$
This nicely joins with the two-parameter bias 
to mean all the usual properties of the ordinary kernel density
estimator with the single exception that the bias depends now on 
$(f-f_0)''$ rather than just $f''$. 

For either three or four parameters, (4.4) yields 
$$\Var\,\hatt f(x)=(nh)^{-1}f(x)
        {\int(k_2z^2-k_4)^2K(z)^2\,\d z\over (k_4-k_2^2)^2}
        -n^{-1}f(x)^2+O(hn^{-1}). \eqno(4.6)$$
And this variance quantity associates appropriately with 
the kernel-dependent quantity given for the three parameters in (4.2)
and for four parameters in (4.3). The two are the bias and variance of the 
fourth order kernel 
$\{(k_2z^2-k_4)/(k_4-k_2^2)\}\,K(z)$; 
see Jones and Foster (1993). This equivalence is familiar for 
local quadratic or cubic regression (Ruppert and Wand, 1994).
But here we observe it for density estimation and, 
most importantly, for {\it any} local three or four parameter model. 

As the pattern is that, for example, five and six parameters 
affords $O(h^6)$ bias, so $O((nh)^{-1})$ variance can be expected,
and an equivalent kernel that is an appropriate quartic multiple of $K$.

We should note briefly that $p$ parameters affords,
again in parallel with $p-1$'th degree polynomial fitting,
natural estimators of the first $p-1$ derivatives of $f$. 
The usual rates for derivative estimation, 
which involves a variance contribution of order $n^{-1}h^{-(2r+1)}$ 
for the $r$'th derivative, can be shown to obtain, 
and equivalent derivative kernels (Ruppert and Wand, 1994) will arise.

\bigskip
{\bf 5. Special cases.}
This section exhibits various special cases of the general
methodology. 

\subsection
{\csc 5.1. The classic kernel method.}
The simplest special case is to set $f(x,\theta)=\theta$.
Semiparametrically, this is not especially attractive since the limiting form
of the estimator as $h\arr\infty$ is uniform 
(albeit an improper uniform). But for small $h$, i.e.~locally
to $x$, this makes perfect sense. 
Moreover, the resulting density estimator is given explicitly by
$$n^{-1}\sumin K_h(x_i-x)\Big/\int K_h(t-x)\,\d t.$$
Since the integral is 1 
the denominator may be ignored, and the
result is precisely the classical kernel density estimator, $\tilda f$.
We mention the denominator, however, because it is not unity near any boundary
of $f$'s support, but rather effects a renormalisation near the boundary
as discussed further in Section 6.1.

Following on from this, a natural first two-parameter locally parametric
estimator is provided by fitting a line, $\theta_1+\theta_2(t-x)$, say,
locally to $x$. Provided we need not worry about boundaries, 
$\int K_h(t-x)(t-x)\,\d t=0$, 
and hence it turns out that $\hatt f(x)=\tilda f(x)$ once more. 
Note that both local constant and linear models have 
$f''_0(x)=0$, and the bias formula (4.1) gives the classic answer 
$\half\sigma_K^2h^2f''(x)$. 
(Near boundaries, local lines automatically adjust $\hatt f$ in a way that
has good consequences which are described in Section 6.2.)

Local polynomials are the obvious further extension, higher degree polynomials
corresponding to higher orders of bias in a way entirely analogous to
local polynomial fitting in regression (e.g.~Ruppert and Wand, 1994).
Local polynomials are not so attractive (in density estimation) in
semiparametric terms, however.

\subsection
{\csc 5.2. Local log-linear density.}
Consider the local model $a\exp(b(t-x))$ for $f$ around $x$
(as does Loader, 1996). 
The score function is $(1/a,t-x)'$, 
and the two equations to solve, in order to 
maximise the local likelihood, are 
$$n^{-1}\sumin K_h(x_i-x)\mtrix{1/a \cr x_i-x \cr}
=\int K_h(t-x)\mtrix{1/a \cr t-x \cr}a\exp(b(t-x))\,\d t. $$
The components on the right hand side can be written 
$\psi(bh)$ and $ah\psi'(bh)$, where $\psi(u)=\int\exp(uz)K(z)\,\d z$
is the moment-generating function for $K$. 
The two equations therefore become $\tilda f(x)=a\psi(bh)$ and 
$\tilda g(x)=ah\psi'(bh)$, where $\tilda g(x)$ is the average 
of $K_h(x_i-x)(x_i-x)$.
Note that the general recipe says 
$\hatt f(x)=f(x,\hatt a(x),\hatt b(x))=\hatt a(x)$, 
so the $\hatt b(x)$ is only somewhat silently present
when using this local reparametrisation. Here one solves 
$\tilda g(x)/\tilda f(x)=h\psi'(bh)/\psi(bh)$ for $b$ and 
in the end uses $\hatt f(x)=\tilda f(x)/\psi(\hatt bh)$. 

This apparatus can be used 
in particular when $K$ is the standard normal.
Some mild caution is called for since $K$ then has unbounded support,
to the effect that the local model is only trusted when 
$t\in x\pm2.5\,h$, say. 
In this case $\tilda g(x)$ above is directly related to the 
derivative $\tilda f'(x)$ of the ordinary kernel estimator,
indeed $\tilda g(x)=h^2\tilda f'(x)$.
(In fact, $\tilda g(x)/(\sigma_K^2h^2)$ 
is quite generally an estimator of $f'$,
usually a different one from $(\tilda f)'$. For comparisons see Jones, 1994.)
This fact, combined with 
$\psi(u)=\exp(\half u^2)$ and $\psi'(u)=\psi(u)u$, gives 
$\hatt b=\tilda f'(x)/\tilda f(x)$ and 
$$\hatt f(x)=\tilda f(x)\exp(-\half h^2\hatt b^2)
        =\tilda f(x)\exp\bigl[-\half h^2
        \{\tilda f'(x)/\tilda f(x)\}^2\bigr]. \eqno(5.1)$$
This particular version of our general local likelihood method 
performs accordingly an explicit correction to the traditional
estimator, attempting to get the local slope right. 
Its bias in general is $\half h^2\{f''-(f')^2/f\}+O(h^4)$, 
which will be only $O(h^4)$ if the true model agrees with 
$a_0\exp(bt)$ on $|t-x|\le2.5\,h$. 

As mentioned in Section 4.3 $\hatt b$ will be more
variable than $\hatt a$, and might require a larger window
parameter for its estimation. The correction factor 
$\hatt b(x)=\tilda f'(x)/\tilda f(x)$ in (5.1) 
could therefore either be computed separately, 
for a somewhat larger $h$ than that used for $\tilda f$,
or the values of $\hatt b(x)$ could be post-smoothed before 
being plugged into (5.1). 

\subsection
{\csc 5.3. Local level, slope, and curvature.}
As a continuation of the previous special case, 
as well as of the theory of Section 4.3 and of Loader (1996), 
one can try out 
$f(t)=a\exp\{b(t-x)+\half c(t-x)^2\}$ for
$t$ in a neighbourhood of $x$. 
This local model is meant to be able to capture local level, 
slope and curvature of the true density, in the neighbourhood 
$t\in x\pm kh$, as above.  
For each given $x$ there are now three equations to solve, 
$$n^{-1}\sumin K_h(x_i-x)\mtrix{1/a \cr x_i-x \cr (x_i-x)^2 \cr}
=\int K_h(t-x)\mtrix{1/a \cr t-x \cr (t-x)^2 \cr}
        ae^{b(t-x)+\half c(t-x)^2}\,\d t. $$
The right hand side gives
three functions in $(a,b,c)$ to equate to 
$\tilda f(x)$, $\tilda g(x)$ (given above) 
and $\tilda g_2(x)=n^{-1}\sumin K_h(x_i-x)(x_i-x)^2$. 
In the end the local likelihood estimator is 
$\hatt f(x)=f(x,\hatt a,\hatt b,\hatt c)=\hatt a$. 

Finite-support kernels are perhaps advisable here,
to secure finiteness of the integrals on the right hand side. 
The equations must in general be solved numerically, for each $x$;
existence and uniqueness of a solution is guaranteed 
by concavity in $(\log a,b,c)$ of the local likelihood.   
But let us give the fairly explicit solution that is possible 
for the case of the standard normal $\phi$ being used
for $K$, interpreting the local model to be an approximation 
on $t\in x\pm2.5\,h$. In this case $\tilda g=h^2\tilda f'$ and 
$\tilda g_2=h^2\tilda f+h^4\tilda f''$, bringing in information about
the first and second derivative of the standard estimator. 
The three equations become
$$\eqalign{
\tilda f(x)&=(a/R)\,\exp(\half h^2b^2/R^2), \cr
\tilda f'(x)&=(ab/R^3)\,\exp(\half h^2b^2/R^2), \cr
h^2\tilda f(x)+h^4\tilda f''(x)&=(ah^2/R^3)(1+h^2b^2/R^2)
        \,\exp(\half h^2b^2/R^2), \cr}$$
where $R=(1-ch^2)^{1/2}$; there is a unique solution 
if the $\hatt c$ found in a minute obeys $1>\hatt ch^2$. 
Some manipulations show that $\hatt R=(1-\hatt ch^2)^{1/2}$ 
can be found from 
$$(1/\hatt R^2)-1=h^2\bigl[\tilda f''(x)/\tilda f(x)
        -\{\tilda f'(x)/\tilda f(x)\}^2\bigr]=h^2\hatt D. $$
This gives $\hatt c=h^2\hatt D/(1+h^2\hatt D)$, and in the end 
$$\hatt f(x)=\tilda f(x)\hatt R\exp\bigl[-\half h^2\hatt R^2
        \{\tilda f'(x)/\tilda f(x)\}^2\bigr]. \eqno(5.2)$$
Note that $\hatt f$ can be computed quite explicitly 
in cases (5.1)--(5.2). This is quite fortunate, of course,
in view of the general complexity of our scheme. 

Again, Loader (1996) has also, independently of the present
authors, worked with local likelihood estimation of 
densities that are log-linear in polynomials. 
Formulae (5.1) and (5.2) are not in Loader (1996),
but he comments further on the general implementation issues involved. 
The manipulations that led to (5.1) and (5.2) do not, 
unfortunately, extend so neatly to the log-cubic case. 

\subsection
{\csc 5.4. A running normal density estimate.}
Let us fit the normal density locally using $1$ and $t-x$ as weight 
functions in (1.4), that is, 
$$n^{-1}\sumin K_h(x_i-x)\mtrix{1 \cr x_i-x \cr}
        =\int K_h(t-x)\mtrix{1 \cr t-x \cr}
        {1 \over \sigma}\phi\Bigl({t-\mu \over \sigma}\Bigr)\,\d t $$
are solved to get hold of the local $\hatt\mu(x)$ and $\hatt\sigma(x)$. 
This should essentially take care of the local level and slope. 
If $K=\phi$ is used, then these equations after some calculations become  
$$\eqalign{
\tilda f(x)
&=\phi\Bigl({x-\mu\over (\sigma^2+h^2)^{1/2}}\Bigr)
        {1\over (\sigma^2+h^2)^{1/2}}, \cr
\tilda f'(x)
&=-{x-\mu\over (\sigma^2+h^2)^{3/2}}
        \phi\Bigl({x-\mu\over (\sigma^2+h^2)^{1/2}}\Bigr), \cr}\eqno(5.3)$$
essentially matching traditional estimates of $f$ and $f'$ 
with quantities predicted by the model. It follows that 
$\tilda q(x)=\tilda f'(x)/\tilda f(x)=-(x-\mu)/(\sigma^2+h^2)$, 
and when inserted in the first equation this gives a single equation to
solve for the local $\sigma=\hatt\sigma(x)$, 
$${1\over \sqrt{2\pi}}{1\over \sqrt{\sigma^2+h^2}}
        \exp\{-\half\tilda q(x)^2(\sigma^2+h^2)\}
        =\tilda f(x). $$
There is a unique solution provided only 
$\phi(h\tilda q(x)) > h\tilda f(x)$. 
Then the local $\mu=\hatt\mu(x)$ is found from 
$\hatt\mu(x)=x+\{\hatt\sigma(x)^2+h^2\}\tilda q(x)$. 

One may alternatively use the local likelihood function (1.3),
i.e.~minimise 
$$n^{-1}\sumin K_h(x_i-x)\{\log\sigma+\half(x_i-\mu)^2/\sigma^2\}
        +\phi\Bigl({x-\mu\over (\sigma^2+h^2)^{1/2}}\Bigr)
        {1\over (\sigma^2+h^2)^{1/2}} \eqno(5.4)$$
to produce $\hatt\mu(x)$ and $\hatt\sigma(x)$. This can be thrown
to an optimiser, or one could use say Newton--Raphson to 
solve the two equations that use the score functions 
$\sigma^{-2}(t-\mu)$ and $\sigma^{-1}\{(t-\mu)^2/\sigma^2-1\}$
as weight functions. These equations can be worked out to be 
$$\eqalign{
n^{-1}\sumin K_h(x_i-x){x_i-\mu\over \sigma}
        &={\sigma(x-\mu)\over (\sigma^2+h^2)^{3/2}}
\phi\Bigl({x-\mu\over \sqrt{\sigma^2+h^2}}\Bigr), \cr
n^{-1}\sumin K_h(x_i-x)\Bigl({(x_i-\mu)^2\over \sigma^2}-1\Bigr) 
        &={\sigma^2\over (\sigma^2+h^2)^{3/2}}
        \phi\Bigl({x-\mu\over \sqrt{\sigma^2+h^2}}\Bigr) \cr
&\qquad\qquad   \Bigl({(x-\mu)^2\over \sigma^2+h^2}-1\Bigr). \cr}$$
The running parameter estimates, for both versions (5.3) and (5.4), 
would now have to be computed over a grid of $x$ values. 
A practical suggestion would be to start 
optimising or equation solving at a new $x$ at the 
optimised values for the previous~$x$. 

The local log-likelihood $L_n(x,\theta)$ is not necessarily concave, 
but it should be so with high probability since the matrix of second 
derivatives goes to the $-J_h$ matrix, defined in Section 3,
and the $J$ matrix is symmetric and positive definite in this $v=u$ case. 
We are hopeful that simplistic computational 
schemes should work well, a problem currently under 
investigation by J.~Fosen, a student of the first author. 

\def\finit{f_{\rm init}}

\subsection
{\csc 5.5. Correcting a parametric start.}
An alternative approach to semiparametric estimation might be to start
with a known, or globally estimated parametric, model $\finit(t)$ and to
multiply it with a local correction factor. Estimation of the local
correction factor can conveniently 
take place within our local likelihood framework as follows. 
First, let $f(t,\theta)=\finit(t)\theta$. 
We think of $\theta=\theta(x)$ as the local correction factor 
for $t$ near $x$. The local log-likelihood is
$\tilda{f}(x)\log\theta - \theta\int K_h(t-x)\finit(t)\,\d t$. 
The resulting estimator is
$$\hatt f(x)=\finit(x)\,{\tilda{f}(x)\over (K_h*\finit)(x)} 
        =\tilda f(x)\,{\finit(x)\over (K_h*\finit)(x)}, \eqno(5.5)$$
where `$*$' denotes convolution. 
Note the simplicity and explicitness of this solution.
The two expressions are meant to make clear 
two useful viewpoints; the estimator is a (typically parametric) 
start estimator times a nonparametric correction, and also
the nonparametric kernel estimator times a parametric update.  

We could also try $f(t)=\finit(t)\,a\exp(b(t-x))$ for $t$ near $x$. 
The local log-likelihood becomes 
$$\log a\,\tilda f(x)+b\tilda g(x)-a\int K(z)\finit(x+hz)\exp(bhz)\,\d z, $$
where $\tilda{g}$ is as before. 
Note that the log-likelihood is concave
in $(\log a,b)$. Maximising the local likelihood 
gives two equations which will not be solvable explicitly in general. 
However, for the normal case and with a normal kernel we find 
$$\hatt f(x)=\tilda f(x)(1+h^2/\sigma^2)^{1/2}
        \exp\bigl[-\half h^2(1+h^2/\sigma^2)
        \{\tilda f'(x)/\tilda f(x)\}^2\bigr]. \eqno(5.6) $$
This is a (simpler) close relation of (5.2). In a way, however, the normal
case is misleading in its potential: formulas like (5.2) and (5.5) are
utilising special properties of the normal to approximate the obvious bias
correction $\tilda{f}(x) - \frac12 h^2 \tilda{f''}(x) $ where $\tilda{f''}$
is an appropriate estimator of $f''$.

Asymptotic bias properties of the above are interesting. 
Both have $b(x)$ of the form $(f-f_0)''(x)$ 
since when the local correction is a constant, 
$u_0$ is a constant also. 
In the local constant correction case, 
$b$ can be written $f''-f \finit''/\finit$; 
in the local exponential-of-linear correction case, 
some further analysis shows that 
the $b$ function can be written 
$f''-(f')^2/f+f(\finit')^2/\finit^2-f\finit''/\finit$. 
Each is zero if $\finit=f$. 
Another appealing $b$ function which can be reached within
this correction factor framework is $b=\finit(f/\finit)''$, 
which is the bias factor function for Hjort and Glad's (1995) estimator. 
These authors' nonparametric correction to a
parametric start arises if the kernel $K(z)$ is replaced by 
the modified (local) kernel $K(z)\finit(x)/\finit(x+hz)$ 
in the local constant correction described above.

An interesting feature of semiparametric estimators 
of the form $\tilda f$ times a parametric correction, 
as the second expression in (5.5), 
is that if taking a likelihood approach,
one need not localise the likelihood but may use 
a global likelihood to estimate the parameters in the parametric part, 
the localisation already being attended to by $\tilda f$. 
Efron and Tibshirani (1996) develop such an approach.
Finally we point out that these 
local nonparametric multiplicative correction methods 
also work well when the initial estimator is itself nonparametric.
When $\finit$ is the kernel method, for example,
(5.5) gives $\tilda f_K^2/\tilda f_{K*K}$, 
subscripts indicating the kernel functions used. Such estimators 
have bias of order $h^4$ and performance generally similar to 
that of an estimator investigated in Jones, Linton and Nielsen (1995);
see Hjort (1996b). 

\subsection
{\csc 5.6. Local $L_2$-fitting.}
Consider the local distance measure 
$\int\! K_h(t-x)\{f(t)\allowbreak-f(t,\theta)\}^2\,\d t$,  
an alternative to the local Kullback--Leibler distance (2.1).
Multiplying out and disregarding the one term which does not 
depend on the parameter we arrive at the following natural
proposal: minimise, for each local $x$, the criterion function 
$$Q_n(x,\theta)=\int K_h(t-x)f(t,\theta)^2\,\d t
        -2n^{-1}\sumin K_h(x_i-x)f(x_i,\theta), $$
and use the accompanying version of $f(x,\hatt\theta(x))$. 
This would constitute a third possible avenue for computing
a running normal estimate, for example. Taking the derivative
it is seen that this local $L_2$-method is a special case
of the general (1.4) method, with weight function 
$v(t,\theta)=f(t,\theta)u(t,\theta)$.
Thus the theory developed applies to this case, 
and suggests in particular that 
the behaviour would be quite comparable to that of the other
methods, for small bandwidths.  
We would prefer the local likelihood to
the local integrated quadratic for large and moderate $h$,
that is, in situations where the parametric model used is not
entirely inadequate, since the likelihood method is more efficient then. 
In the normal case, if $h$ is large, 
the variance of the $\mu$ estimator is about 1.54 times higher 
with the $L_2$ method and the variance
of the $\sigma$ estimator about 1.85 times higher.  
However, the corresponding parameter estimates are more robust
than the maximum likelihood ones. 
Further results and discussion are in Hjort (1994). 

\subsection
{\csc 5.7. Uniform kernel.} 
Let $K$ be uniform on $[-\half,\half]$. In this case 
the local log-likelihood function is 
$L_n(x,\theta)=n^{-1}\sum_W\log f(x_i,\theta)
        -\{F(x+\half h,\theta)-F(x-\half h,\theta)\}$,
where the sum is over the window where $x_i\in x\pm\half h$.
Maximising this essentially aims to match empirical facts 
from the local window $x\pm\half h$ to behaviour predicted 
by the parametric $f(\cdot,\theta)$ on this window. If $v_1(x,t,\theta)=1$
is one of the weights used in (1.4) then that equation simply 
matches the empirical and theoretical probabilities of falling 
inside this window. 

\subsection 
{\csc 5.8. Relationship with moment estimation.}
Note that as $h$ becomes large, the (1.4) recipe ends up 
choosing as estimate the parameter value that solves 
$n^{-1}\sumin v(x_i,\theta)=\E_\theta v(X_i,\theta)$,
which is ordinary moment estimation with the $v_j(X_i,\theta)$ 
functions. This also indicates that having $v_1(t,\theta)=1$ as first 
weight function, which we partly used in special cases above, 
does not work well with large $h$s. 
We would expect the two methods of obtaining a running 
normal density estimate, based on equations (5.3) and (5.4) respectively, 
to perform similarly for small $h$s, 
but the second method would perhaps be the best one 
for moderate and large $h$s. 

\bigskip
{\bf 6. Estimating the density at a boundary.} 
Throughout the theoretical exposition so far,
we have assumed that $f$
has support the whole real line. In this section, we consider the presence
of known boundaries to $f$'s support. It will be general enough to consider
positive data, and hence one boundary at zero. Consider estimation points $x$
at and near the boundary in the sense that $x = ph$ for $0 \leq p <1$ and
suppose $K$ has support $[-1,1]$. (This setup can easily be extended to
infinite support kernels but is standard and delineates boundary and interior
regions, results already proved continuing to hold for $x$ in the interior.)
Define
$a_l(p) = \int_{-1}^p u^l K(u)\,\d u$
and
$b(p) = \int_{-1}^p K^2(u)\,\d u$.
(Note that for $p \geq 1$, 
$a_0(p) = 1$, $a_1(p) = 0$, 
$a_2(p) = \sigma_K^2$
and $b(p) = R(K)$.)

\subsection
{\csc 6.1. The one-parameter case.}
For $x$ near the boundary, formula (3.4) changes to
$$\int K_h(t-x)g(t)\,\d t=a_0(p)g(x)- a_1(p) h g'(x) + 
        \half a_2(p)h^2g''(x)+O(h^3). \eqno(6.1)$$ From this, 
it immediately follows that
$$\E\hatt f(x) \simeq f(x) - \{a_1(p)/a_0(p)\}h(f-f_0)'(x). $$
With a single locally fit parameter, therefore, 
boundary bias is of the undesirable $O(h)$ type unless one has been
fortunate enough to choose one's parametric class equal to the true $f$
near the boundary.
The boundary variance follows from (4.4) when $V_h(z)=1$: 
$$J=a_0(p)f_0(x)+O(h) \quad {\rm and} \quad 
M=b_0(p)f(x)+O(h). \eqno(6.2)$$
These give a variance of
$$\Var\,\hatt f(x) \simeq (nh)^{-1} \{b(p)/a_0^2(p)\} f(x) . \eqno(6.3)$$

Bias and variance in (6.1) and (6.3) exactly match those of
a standard kernel estimator divided by $a_0(p)$ save the replacement of $f'$
by $f'-f_0'$ (e.g.~Jones, 1993b). That is,
the one parameter local likelihood estimator behaves
much like a renormalised kernel estimator in respect of boundaries.
In Section 5.1 we noted that if the single parameter were a constant, such a
renormalisation explicitly and exactly takes place; however, the
current asymptotic observations apply
more generally to any one-parameter fitting.

\subsection
{\csc 6.2. The two-parameter case.}
Just as the local linear regression fit has an appealing $O(h^2)$ boundary
bias (Fan and Gijbels, 1992), so too does the two parameter locally parametric
density estimator, as we shall now demonstrate. 

Write $v_{0,j}$ for the derivative of $v$ with respect to $\theta_j$, $j=1,2,$
evaluated at $\theta_0$. 
To obtain the bias, we need to study the expansions of
$$\int_0^\infty K_h(t-x) v_{0,1} (t) \{f(t)-f(t_0) \}\,\d t=0
= \int_0^\infty K_h(t-x) v_{0,2} (t) \{f(t)-f(t_0) \}\,\d t. $$
Expanding each to order $h^3$, and writing $(f-f_0) \simeq Ah^2$, 
$(f-f_0)' \simeq Bh$ and $(f-f_0)'' \simeq C$, we find that the $O(h^2)$
term in either side of the above expression involves 
$A-\{a_1(p)/a_0(p)\}B+\half\{a_2(p)/a_0(p)\}C$ 
and that the difference between left- and right-hand sides 
yields an $O(h^3)$ term involving $-a_1(p)A +a_2(p)B-\half a_3(p)C$.
Setting these two quantities to zero and solving for $A$ yields
$$\E\hatt f(x) = f(x)+\half Q(p) h^2 \{f''(x) - f_0''(x)\}, \eqno(6.4)$$
where
$$Q(p) = {a_2^2(p) - a_1(p) a_3(p) \over a_2(p) a_0(p) - a_1^2 (p)} .$$

For the variance in the two-parameter case, 
simply use $V_h(z)=(1,hz)$ in (4.4); we get 
$$\Var\,\hatt f(x)=(nh)^{-1}f(x)
{\int\{a_2(p)-a_1(p)z\}^2k(z)^2\,\d z
        \over \{a_0(p)a_2(p)-a_1(p)^2\}^2}+O(n^{-1}). \eqno(6.5)$$
The kernel-dependent asymptotic bias and variance terms 
are precisely those of the popular boundary kernel 
$${a_2(p)-a_1(p)z \over a_0(p)a_2(p)-a_1(p)^2}K(z) $$
(see e.g.~Jones, 1993b). That is, with two parameters we 
achieve $O(h^2)$ boundary bias (regardless of choice of local model)
in an appealing way, and there is also the potential of 
further decrease in bias due to a good choice of model. 

Three parameters can be expected to achieve $O(h^3)$ boundary bias,
four parameters $O(h^4)$, but this is not pursued here. 

\bigskip
{\bf 7. Multi-dimensional data.} 
The local likelihood method based on (1.3) generalises easily 
to the case of $d$-dimensional data vectors,
using $d$-dimensional kernel functions.  
The general weight function version of the method, 
through solving (1.4) for as many equations as there are parameters
in the model used, is also operable in the vector case. 
A bivariate example could be to smooth the product-normal model, 
where the final estimator is of the form
$$f(x,y,\hatt\mu_1,\hatt\mu_2,\hatt\sigma_1,\hatt\sigma_2)
=\hatt\sigma_1^{-1}\phi(\hatt\sigma_1^{-1}(x-\hatt\mu_1))\,
        \hatt\sigma_2^{-1}\phi(\hatt\sigma_2^{-1}(x-\hatt\mu_2)). $$
This would smooth towards normal marginals but also smooth 
somewhat towards independence. 

Defining such estimators is therefore easy in principle,
although computational matters become more complicated with
the increasing number of running parameters to solve for. 
The local minimum Kullback--Leibler distance result of 
Section 2.1 is also seen to hold, giving support to the idea.  
Another question is to what extent the theory of the
previous sections can be generalised, to establish 
properties of the resulting density estimators. 
We shall briefly go through the two-dimensional case to illustrate
that the theory indeed goes through with appropriate 
extensions of previous techniques.
Again it will be seen that the new method has 
scope for reduction of bias in a large neighbourhood 
of densities around the parametric model employed. 
Our machinery could perhaps turn out to be of particular value
in the multi-dimensional case, where there is much to lose and 
appalling convergence rates to meet by not imposing any structure at all. 
 
Let $K(z_1,z_2)=K_1(z_1)K_2(z_2)$ be a product kernel. 
A good version of the traditional estimator is 
$$\tilda f(x_1,x_2)=n^{-1}\sumin K_{h_1,h_2}(x_{i,1}-x_1,x_{i,2}-x_2)
        =n^{-1}\sumin K_{h_1,h_2}({\bf x}_i-{\bf x}), $$
where $K_{h_1,h_2}({\bf t})=h_1^{-1}K_1(h_1^{-1}t_1)h_2^{-1}K_2(h_2^{-1}t_2)$,
and where we write $\bfx=(x_1,x_2)$ and so on; 
see Wand and Jones (1993). It has 
$${\rm bias}\simeq
\sum_{i=1}^2\half\sigma(K_i)^2h_i^2f''_{ii}(\bfx)
{\rm\ and\ }
{\rm variance}\simeq{R(K_1)R(K_2)f(\bfx)\over nh_1h_2}
        -{f(\bfx)^2\over n}, \eqno(7.1)$$ 
where $\sigma(K_i)^2=\int z^2K_i(z)\,\d z$ 
and $R(K_i)=\int K_i(z)^2\,\d z$. 
We also use $f''_{ii}(\bfx)$ for $\dell^2f(\bfx)/\dell x_i^2$ and so on. 

The new locally parametric estimator is defined as 
$\hatt f(\bfx)=f(\bfx,\hatt\theta(\bfx))$, 
where the local parameter estimate solves 
$$n^{-1}\sumin K_{h_1,h_2}(\bfx_i-\bfx)v(\bfx_i,\theta) 
        -\int K_{h_1,h_2}(\bft-\bfx)v(\bft,\theta)f(\bft,\theta)
        \,\d\bft={\bf 0} $$
around each given $\bfx$ point. Four independent equations are
needed to handle the product-normal model above, for example. 
The expected value of $\hatt f(\bfx)$ is $f(\bfx,\theta_0)+O((nh)^{-1})$, 
where $\theta_0$ is locally least false and solves
$$V_j(\bfx,\theta)=\int K_{h_1,h_2}(\bft-\bfx)v_j(\bft,\theta)
        \{f(\bft)-f(\bft,\theta)\}\,\d\bft=0 
        \quad{\rm for\ }j=1,\ldots,p.$$
Using           
$$\int K_{h_1,h_2}(\bft-\bfx)g(\bft)\,\d\bft
=g(\bfx)+\sum_{i=1}^2\half\sigma(K_i)^2h_i^2g''_{ii}(\bfx)
        +O((h_1^2+h_2^2)^2),$$
which is proved by Taylor expansions and properly 
generalises (3.4), one finds that 
$${\rm bias}
\simeq\sum_{i=1}^2\half\sigma(K_i)^2h_i^2
\Bigl[f''_{ii}(\bfx)-f''_{0,ii}(\bfx)
        +2{v'_{j,0,i}(\bfx)\over v_{j,0}(\bfx)}
        \{f'_i(\bfx)-f'_{0,i}(\bfx)\}\Bigr], $$
where $f_0$ and $v_0$ indicate the $f(\bft,\theta)$ and $v(\bft,\theta)$
functions with $\theta_0=\theta_0(\bfx)$ inserted. 
If there is more than one $v_j$ function in direction $x_i$, then 
$f'_i(\bfx)-f'_{0,i}(\bfx)$ is necessarily $o(1)$, and 
$${\rm bias}\simeq
\sum_{i=1}^2\half\sigma(K_i)^2h_i^2
        \{f''_{ii}(\bfx)-f''_{0,ii}(\bfx)\}. \eqno(7.2)$$
Further, even $f''_{ii}(\bfx)-f''_{0,ii}(\bfx)$ is $o(1)$ in directions
involving three of more $v_j$ functions and bias order 
can then be reduced, and so on.
Turning next to the variance, one needs to consider
$$M=h_1h_2\,{\rm VAR}_f\{K_{h_1,h_2}(\bfX_i-\bfx)v_0(\bfX_i)\} $$
and 
$$J=\int K_{h_1,h_2}(\bft-\bfx)\bigl[v_0(\bft) 
        u_0(\bft)'f_0(\bft)+v_0^*(\bft)\{f_0(\bft)
        -f(\bft)\}\bigr]\d\bft. $$
Using the same type of method as that used in Sections 4.2--4.3 
in this more laborious situation 
one ends up with exactly the same variance as in (7.1), 
to the order of approximation used,
provided there are no more than two local parameters 
in each direction. (Extensions to higher numbers of 
parameters can be carried out, as with the case that led to equation (4.4).)

An interesting special case of the general method is that of 
a local model $f(t_1,t_2)=a\exp(b_1(t_1-x_1)+b_2(t_2-x_2))$,
for $\bft$ around $\bfx$, modelling local level and local slopes. 
The score function is $(1/a,t_1-x_1,t_2-x_2)'$, and gives three 
equations to solve for the three parameters. If the product
normal kernel is used calculations generalising those of Section 5.2
yield 
$$\hatt f(\bfx)=\tilda f(\bfx)\exp\Bigl[-\half\sum_{i=1}^2
        h_i^2\{\tilda f'_i(\bfx)/\tilda f(\bfx)\}^2\Bigr]. \eqno(7.3)$$
A more involved version can be given where the local 
curvatures $\exp\{\half c_i(t_i-x_i)^2\}$ and/or the 
local covariance factor $\exp\{d(t_1-x_1)(t_2-x_2)\}$ are taken into account,
thus generalising the one-dimensional (5.2).
Yet another estimator of interest evolves by modelling 
$f(\bft)$ as a global $\finit(\bft)$ times a local 
log-linear correction factor, in the spirit of Section 5.5. 
Explicit estimators can be written out, similar to formula (5.6),
for the case of a binormal start and Gau\ss ian kernels. 

\bigskip
{\bf 8. Supplementing results and remarks.}  

\subsection
{\csc 8.1. mse and mise analysis.}
The approximate mean squared error 
for the new estimator is 
$${\rm amse}\{\hatt f(x)\}=\quart\sigma_K^4h^4b(x)^2+R(K)(nh)^{-1}f(x), $$
with $b(x)=f''(x)-f_0''(x)$ in the typical case,
and ignoring terms of order $n^{-1}+h^6+h/n$ or smaller. 
For estimation consistency we need $h\arr0$ (forcing the bias 
to zero) while $nh\arr\infty$ (forcing variance to zero). 
The theoretically best choice of $h$ at $x$ is therefore of the form 
$\{R(K)/\sigma_K^4\}^{1/5}\{f(x)/b(x)^2\}^{1/5}\allowbreak\,n^{-1/5}$,
and the theoretically best amse is 
${5\over 4}\{R(K)\sigma_K\}^{4/5}f(x)^{4/5}b(x)^{2/5}\,n^{-4/5}$. 
Choosing the best $h$ for every $x$ is generally too ambitious,
and it is convenient to study the approximate or asymptotic 
mean integrated squared error 
${\rm amise}(\hatt f)=\quart\sigma_K^4h^4R_{\rm new}(f)
        +R(K)(nh)^{-1}$, 
where $R_{\rm new}(f)=\int b(x)^2\,\d x$. The theoretically 
best global $h$-value is 
$$h_0=\{R(K)/\sigma_K^4\}^{1/5}R_{\rm new}(f)^{-1/5}\,n^{-1/5}, \eqno(8.1)$$
leading to the theoretically best amise 
${5\over 4}\{R(K)\sigma_K\}^{4/5}R_{\rm new}^{1/5}(f)\,n^{-4/5}$. 
We note that the Yepanechnikov kernel $K_0(z)={3\over2}(1-4z^2)_+$
(and scaled versions thereof) 
is optimal in that it manages to minimise $R(K)\sigma_K$, 
see e.g.~Wand and Jones (1995, Section 2.7). 

\subsection
{\csc 8.2. Comparison with the traditional method.} 
The calculations above are quite analogous to well known ones for the 
ordinary kernel method (which in any case are a special case).
This also makes it easy to compare the two
methods. Using the global (approximate) mise criterion 
we see that the new method is better provided 
$R_{\rm new}(f)<R_{\rm trad}(f)$, 
where the latter roughness quantity is $\int (f'')^2\,\d x$. 
This statement refers to the situation where both methods use
the same kernel and the same bandwidth. If $R_{\rm new}$ really
is smaller, then $\hatt f$ can be made even better by selecting a 
better $h$. This also defines a relatively 
broad nonparametric neighbourhood of densities around the parametric
model at which the new method is better. 
That $R_{\rm new}$ really offers a significant improvement 
on $R_{\rm trad}$ in many practical situations, 
for some of the new estimators displayed in Section 5, 
will be substantiated and reported on in future work. 

At a pointwise level, several points made by Hjort (1996a) 
in the analogous locally parametric hazard estimation case are worth
repeating, in modified form, here.
First, it is easy to show that the locally parametric estimator is
(asymptotically) better than the classical estimator whenever
$0\le f''_0(x)/f''(x)\le 2$. 
As long as $f''_0$ and $f''$ have the same sign, 
$|f''_0(x)|$ can afford to range over $[0, 2|f''(x)|]$. 
Note that this observation holds for small $h$ i.e.\ at
`the nonparametric end' of our semiparametric estimator. 
We should also note, however, that differences in the constant 
involved in the bias may not be all that important, 
since the squared bias makes up only $1/5$ of optimised
mean squared error, the remainder being due to variance.

The locally parametric estimator is also designed to have especial advantages
over the kernel estimator when $f$ is, in fact, close to $f_0$. Regardless
of this, the kernel estimator has mean squared error of order $h^4 + (nh)^{-1}$
which is minimised by taking $h \sim n^{-1/5}$ and hence optimal mean squared
error of $O(n^{-4/5})$. On the other hand, one might quantify closeness of
$f_0$ and $f$ by setting 
$(f_0-f)''\sim n^{-\epsilon}$ for some $0<\epsilon<\half$. 
The mean squared error of the locally parametric estimator is thus
$h^4 n^{-2 \epsilon } + (nh)^{-1}$ which is optimised by taking 
$h\sim n^{-(1-2 \epsilon)/5}$. 
The optimised mean squared error is then of order $n^{-(4+2\epsilon )/5}$. 
For instance, if $f''_0$ and $f_0$ are $O(n^{-1/4})$ apart, 
the mean squared error is improved to $O(n^{-9/10})$, 
and as the difference tends to $n^{-1/2}$, 
the mean squared error tends to $n^{-1}.$

\subsection
{\csc 8.3. Choosing the bandwidth.} 
Methods for automatic bandwidth selection for the traditional kernel density
estimator are reviewed by Jones, Marron and Sheather (1995). They might
be utilised unaltered for locally parametric estimates, at least as a first
attempt. However, if we are using an estimator that does indeed improve on
the basic one, we will be oversmoothing relative to the new optimal choice.
An argument in Section 8.2 suggests that the degree of oversmoothing 
may not often be very great, however.

The best of the bandwidth selectors in the ordinary case are founded on good
estimates of unknown quantities in mise expressions. The key is usually in
the estimation of $R_{\rm trad} = R(f'')$, and this transfers to the need
to estimate $R_{\rm new} = R((f-f_0)'')$
(one might think of adapting traditional selectors by multiplying them by 
an estimate of $(R_{\rm trad}/R_{\rm new})^{1/5}$). 
But the estimation of $R_{\rm new}$
is not straightforward since it involves the second derivatives of both the
true $f$ and its best possible approximant of the form $f(x,\theta(x))$.
One {\it possibility} might be to estimate $f''$ by $\hatt f''$
using a bandwidth $g$
optimal for $R(f'')$ (this is what happens in a good bandwidth selector for
the traditional estimator, see e.g.~Sheather and Jones, 1991) 
and $f_0''$ by $\hatt f''$
using the same $h$ as for estimation of $f$.
This type of difficulty extends to rule-of-thumb approaches too.
We should also mention that it could be worthwhile to employ 
more than one bandwidth when forming an estimator based on 
several equations, as for the methods of Sections 5.2 and 5.3. 
This is because local slope equations typically would 
benefit from larger bandwidths than for local level equations. 

Least squares cross-validation, 
which for the traditional estimator is less reliable than
the best methods (Jones, Marron and Sheather, 1995), has the advantage that
it doesn't explicitly involve $f_\theta$. One can just follow the usual
idea of estimating $E\{\int\hatt f(x)^2\,\d x-2\int f(x)\hatt f(x)\,\d x\}$
by $\int f(x,\hatt \theta(x))^2\,\d x
-2n^{-1}\sumin f(x,\hatt \theta_{(i)}(x_i))$ where
numerical integration is used for the first term and $\hatt \theta_{(i)}(x_i)$
is the leave-one-out version of $\hatt \theta $.

Alternative methods are also worth considering,
particularly since one sometimes would be interested 
in using moderate or large $h$s, namely in situations 
where the data fit the local model well.  
A changing and adaptively defined $h$ could be advantageous
in some cases. Hjort (1996a) considers a local goodness of fit 
approach in the hazard case: 
increase the bandwidth until the local model fails to pass a goodness of
fit criterion. Extension of this methodology 
to the density case is an interesting topic for further research,
one possibility being to exploit results of Section 8.5 below.

\subsection
{\csc 8.4. Large-sample normality.} 
The basic bias and variance results for our estimator $f(x,\hatt\theta(x))$ 
were derived in Sections 3 and 4. Our arguments were based on 
claims (3.2) and (3.3) about limiting normality for $\hatt\theta(x)$,
and in fact also on variants of these that work in the 
framework where the smoothing parameter $h$ is not fixed 
but goes to zero with $n$. Here we outline proofs of 
precise versions of these claims.  

The $\hatt\theta(x)$ we consider is the solution to (1.4). 
For convenience we partly suppress the fixed $x$ in the notation now. 
Taylor expansion analysis for $V_n(\hatt\theta)=0$ gives 
$$(nh)^{1/2}(\hatt\theta-\theta_0)\simeq -V_n^*(\theta_0)^{-1}
        (nh)^{1/2}V_n(\theta_0), \eqno(8.2)$$
where $V_n^*$ is the $p\times p$ matrix of partial derivatives
of the $V_{n,j}(\theta)$ functions, and this leads to 
a $J_h^{-1}\normal_p\{0,M_h\}$ limit by well known arguments. 
A more formal proof starts out by observing that 
$\hatt\theta$ can be seen as the functional $T(F_n)$, 
where $T(F)$ is the solution to 
$w(F,\theta)=\int K_h(t-x)v(t,\theta)\{\d F(t)-f(t,\theta)\,\d t\}=0$,
see (3.1). Under regularity assumptions this is a second order 
smooth functional in the sense of Shao (1991), with influence function 
$$I(F,t)=J_h^{-1}\Bigl\{K_h(t-x)v(t,\theta_0)
        -\int K_h(t-x)v(t,\theta_0)f(t,\theta_0)\,\d t\Bigr\}, $$
in which $\theta_0=T(F)$. This is seen from a Taylor expansion
of $v((1-\eps)F+\eps\delta_t,\theta)$ around $\theta_0$,
where $\delta_t$ is unit point mass at $t$. 
This is sufficient for consistency and a normal 
$\{0,J_h^{-1}M_h(J_h')^{-1}\}$ limit for 
$(nh)^{1/2}(\hatt\theta-\theta_0)$, see Shao (1991).
These arguments, in conjunction with the theory and tools
developed in Sections 4.2--4.3, can also be used to prove  
$$(nh)^{1/2}\{f(x,\hatt\theta(x))-f(x)-b_n(x)\}
        \arr_d\normal\{0,\tau(K)^2f(x)\}, \eqno(8.3)$$
when $h\arr0$ and $nh\arr\infty$.
Here $b_n(x)$ is the bias of $f(x,\hatt\theta(x))$, 
and is of the form $\half\sigma_K^2h^2b(x)+o(h^2)$ for appropriate 
$b(x)$ functions in the case of one- and two-parameter local 
families, and of the form $c(K)h^4b(x)+o(h^4)$ for certain other 
$b(x)$ functions in the case of three- and four-parameter local families;
see equations (3.7), (4.1), (4.2) or (4.3). 
Also, $\tau(K)^2$ is the general variance factor 
appearing in equation (4.4). 

Another useful version of such a precise result,
valid in the general log-linear case, 
cf.~the special cases treated in Sections 4.2--4.3 and 5.2--5.3, 
is as follows. Let the model be of the form 
$f(t,\theta)=\exp\{\theta'w(t)\}$, where $w(t)$ is a vector of 
$p$ functionally independent and twice differentiable weight functions. 
We assume that $\theta'w(t)$ spans the full real line as 
$\theta$ varies. The local likelihood 
$$L_n(x,\theta)=n^{-1}\sumin K_h(x_i-x)\theta'w(x_i)
-\int K_h(t-x)\exp\{\theta'w(t)\}\,\d t $$
is concave in $\theta$. Let $\theta_{0,h}$ be the unique maximiser 
of the limit function, or, equivalently, the unique solution to 
$\int K_h(t-x)w(t)[f(t)-\exp\{\theta'w(t)\}]\,\d t=0$. 
Next study the function 
$$A_n(s)=nh\{L_n(x,\theta_{0,h}+s/(nh)^{1/2})-L_n(x,\theta_{0,h})\}. $$
It is concave in $s$, 
and inspection shows that it can be expressed as 
$s'U_n-\half s'J_ns+O(\|s\|^3/(nh)^{1/2})$. Here 
$$U_n=n^{-1/2}\sumin h^{1/2}\{K_h(x_i-x)w(x_i)-\xi_n\}, $$
with $\xi_n=\int K_h(t-x)w(t)f(t)\,\d t$, and 
$J_n=\int K_h(t-x)f_0(t)w(t)w(t)'\,\d t$. 
The point is now that the maximiser of $A_n(s)$, which is 
$(nh)^{1/2}(\hatt\theta-\theta_{0,h})$, must be close to 
the maximiser of the quadratic approximation $s'U_n-\half s'J_ns$,
which is $J_n^{-1}U_n$. 
Precise general concavity-based arguments are in Hjort and Pollard (1996).
Now $J_n^{-1}U_n$ has a covariance matrix which stabilises as 
$n$ grows, and using the Lindeberg theorem 
it is not difficult to show that it is asymptotically normal. 
The delta method, combined with the arguments that led to (3.6) and (4.4),
then gives the appropriate version of (8.3) again. 

\subsection
{\csc 8.5. Parameter inspection.}
Plotting the estimated running parameter $\hatt\theta(x)$ against
$x$ is a natural idea. This could be used for model exploration
purposes and for goodness of fit testing. 
Monitoring $\hatt\theta(x)$ based on a pilot value of $h$ 
can also be used for choosing the final bandwidth, 
or for post-smoothing before being used in the final $f(x,\hatt\theta(x))$. 

From the discussion of Section 3.1 it is clear that $\hatt\theta(x)$ 
aims at the locally least false parameter value $\theta_0(x)$, 
which is a constant value $\theta_0$ independent of $x$ 
if and only if the parametric model used is perfect. 
The approximate precision of $\hatt\theta(x)$ can be worked out from 
$J_h^{-1}M_h(J_h')^{-1}$ of (3.2) using methods developed in connection 
with equations (3.6) and (4.4). 
To a first order approximation the variances 
for the components of $\hatt\theta(x)$ are inversely 
proportional to $nhf(x)$ and hence their plots  
can not normally be trusted in regions of small density. 
We note that both weight functions $v(t,\theta)$ 
as well as characteristics of the model used show up in
explicit calculations for the variance matrix for $\hatt\theta(x)$, 
in contrast with the analogous calculation for the variance of 
$f(x,\hatt\theta(x))$, ending with (3.8) and (4.4).


\bigskip
{\bf 9. Conclusions.}
We believe we have been studying {\it the} most attractive way of doing
semiparametric density estimation. The estimators run the gamut from a
fully parametric fit to almost fully nonparametric (except with some small
change in performance which may well be beneficial) with only a single
smoothing parameter to be chosen. The number of parameters in the `local
model' crucially affects performance: one and two fitted parameters are
most readily comparable with ordinary kernel density estimation, three and
four fitted parameters with fourth order kernel estimation, and more
parameters with higher order estimates. Even numbers of fitted parameters
have advantages in terms of simplicity and interpretability of bias. These
comments parallel the fitting of local polynomials in regression, but we note
that they are driven by numbers of parameters only (which are effectively
automatically reparametrised into intercept, slope, curvature, etc.,
parameters) and not by the specific functional form. Together with, and in
generalisation of, Loader (1996), we believe we have laid firm theoretical
foundations for locally parametric nonparametric density estimation. 
Much still remains to be done in terms of exploring practical issues and
applications.

\bigskip
{\bf Acknowledgements.}
We have had fruitful discussions with  
John Copas, Oliver Linton and Jens Perch Nielsen.
Comments from referees inspired improvements over an earlier version. 
Part of this work was carried out while the second author visited 
the University of Oslo with partial support from 
its Department of Mathematics. 

\bigskip
\bigskip
 
\parindent0pt
\parskip3pt
\baselineskip11pt

\def\ref#1{{\noindent\hangafter=1\hangindent=20pt
  #1\smallskip}}  
  
\centerline{\bf References}

\medskip



\ref{%
Buckland, S.T. (1992). 
Maximum likelihood fitting of Hermite and simple polynomial densities.
{\sl Applied Statistics} {\bf 41}, 241--266.} 

\ref{%
Cleveland, W.S. (1979).
Robust locally weighted regression and smoothing scatterplots.
{\sl Journal of the American Statistical Association} {\bf 74}, 829--836.}

\ref{%
Copas, J.B. (1995). 
Local likelihood based on kernel censoring.
{\sl Journal of the Royal Statistical Society} Series B {\bf 57}, 
221--235.} 

\ref{%
Efron, B.~and Tibshirani, R. (1996).
Using specially designed exponential families for density estimation.
To appear.}

\ref{%
Fan, J.~(1992).
Design-adaptive nonparametric regression.
{\sl Journal of the American Statistical Association} {\bf 87},
998--1004.}

\ref{%
Fan, J.~(1993).
Local linear regression smoothers and their minimax efficiencies.
{\sl Annals of Statistics} {\bf 21}, 196--216.}

\ref{%
Fan, J.~and Gijbels, I. (1992).
Variable bandwidth and local linear regression smoothers.
{\sl Annals of Statistics} {\bf 20}, 2008--2036.}

\ref{%
Fan, J.~and Gijbels, I. (1996).
{\sl Local Polynomial Modelling and its Applications.}
Chapman and Hall, London.} 

\ref{%
Fan, J., Heckman, N.E. and Wand, M.P. (1995). Local polynomial kernel
regression for generalized linear models and quasi-likelihood functions.
{\sl Journal of the American Statistical Association} {\bf 90},
141--150.} 

\ref{%
Fenstad, G.U.~and Hjort, N.L. (1996).
Two Hermite expansion density estimators,
and a comparison with the kernel method. 
Manuscript.} 

\ref{%
Hastie, T.~and Loader, C.R. (1993).
Local regression: automatic kernel carpentry (with comments).
{\sl Statistical Science} {\bf 8}, 120--143.} 

\ref{%
Hastie, T.~and Tibshirani, R. (1990).
{\sl Generalized Additive Models.} Chapman and Hall, London.}

\ref{%
Hjort, N.L. (1986).
{\sl Theory of Statistical Symbol Recognition.}
Research Monograph, Norwegian Computing Centre, Oslo.}

\ref{%
Hjort, N.L. (1991).
Semiparametric estimation of parametric hazard rates.
In {\sl Survival Analysis: State of the Art} (eds.~P.S. Goel and J.P.~Klein)
Kluwer, Dordrecht, pp.~211--236. }

\ref{%
Hjort, N.L. (1994).
Minimum L2 and robust Kullback--Leibler estimation.
Proceedings of the {\sl 12th Prague Conference 
on Information Theory, Statistical Decision Functions and Random Processes},
102--105. }

\ref{%
Hjort, N.L. (1995).
Bayesian approaches to semiparametric density estimation.
{\sl Bayesian Statistics V} 
(J.~Bernardo, J.~Berger, P.~Dawid, A.F.M.~Smith, eds.). 
Oxford University Press.}

\ref{%
Hjort, N.L. (1996a). 
Dynamic likelihood hazard estimation.
{\sl Biometrika}, to appear.}

\ref{%
Hjort, N.L. (1996b).
Multiplicative higher order bias kernel density estimators.
Statistical research report, Department of Mathematics,
University of Oslo.} 

\ref{%
Hjort, N.L.~and Glad, I.K. (1995).
Nonparametric density estimation with a parametric start.
{\sl Annals of Statistics} {\bf 23}, 882--904.} 

\ref{%
Hjort, N.L.~and Pollard, D.B.~(1996).
Asymptotics for minimisers of convex processes.
{\sl Annals of Statistics}, to appear.}

\ref{%
Jones, M.C. (1993a).
Kernel density estimation when the bandwidth is large.
{\sl Australian Journal of Statistics} {\bf 35}, 319--326.}

\ref{%
Jones, M.C. (1993b).
Simple boundary correction for kernel density estimation.
{\sl Statistics and Computing} {\bf 3}, 135--146.}

\ref{%
Jones, M.C. (1994).
On kernel density derivative estimation.
{\sl Communications in Statistics --- Theory and Methods.} {\bf 23},
2133--2139.}

\ref{%
Jones, M.C. (1995).
On close relations of local likelihood density estimation. To appear.}

\ref{%
Jones, M.C., Davies, S.J. and Park, B.U. (1994).
Versions of kernel-type regression estimators. 
{\sl Journal of the American Statistical Association} {\bf 89}, 825--832.}

\ref{%
Jones, M.C.~and Foster, P.J. (1993).
Generalized jackknifing and higher order kernels.
{\sl Journal of Nonparametric Statistics} {\bf 3}, 81--94.}

\ref{%
Jones, M.C.~and Hjort, N.L. (1994).
Local fitting of regression models by likelihood: what's important?
Statistical Research Report, Department of Mathematics, 
University of Oslo.} 

\ref{%
Jones, M.C., Marron, J.S. and Sheather, S.J. (1995).
A brief survey of bandwidth selection for density estimation.
{\sl Journal of the American Statistical Association}, top appear.}

\ref{%
Jones, M.C., Linton, O.~and Nielsen, J.P. (1995).
A simple and effective bias reduction method 
for density and regression estimation.
{\sl Biometrika} {\bf 82}, 327--338.}

\ref{%
Lindsey, J.K. (1974).
Comparison of probability distributions.
{\sl Journal of the Royal Statistical Society Series B} {\bf 36}, 38--47.}

\ref{%
Loader, C.R. (1996). 
Local likelihood density estimation. 
{\sl Annals of Statistics}, to appear.} 

\ref{%
Olkin, I.~and Spiegelman, C.H. (1987).
A semiparametric approach to density estimation.
{\sl Journal of the American Statistical Association} {\bf 82}, 858--865.}

\ref{%
Ruppert, D.~and Wand, M.P. (1994).
Multivariate locally weighted least squares regression.
{\sl Annals of Statistics} {\bf 22}, 1346--1370.} 

\ref{%
Schuster, E. and Yakowitz, S. (1985).
Parametric/nonparametric mixture density estimation with application to
flood-frequency analysis.
{\sl Water Resources Bulletin} {\bf 21}, 797--804.}

\ref{%
Scott, D.W. (1992).
{\sl Multivariate Density Estimation:
Theory, Practice, and Visualization.}
Wiley, New York.}

\ref{%
Shao, J. (1991).
Second-order differentiability and jackknife.
{\sl Statistica Sinica} {\bf 1}, 185--202.}

\ref{%
Staniswalis, J. (1989).
The kernel estimate of a regression function in likelihood-based models.
{\sl Journal of the American Statistical Association} {\bf 84}, 276--283.}

\ref{%
Stone, C.J. (1977).
Consistent nonparametric regression.
{\sl Annals of Statistics} {\bf 5}, 595--620.}

\ref{%
Tibshirani, R.~and Hastie, T. (1987).
Local likelihood estimation. 
{\sl Journal of the American Statistical Association} {\bf 82}, 559--567.}

\ref{%
Wand, M.P.~and Jones, M.C. (1993).
Comparison of smoothing parameterizations in bivariate kernel 
density estimation.
{\sl Journal of the American Statistical Association} {\bf 88}, 520--528.}

\ref{%
Wand, M.P.~and Jones, M.C. (1995).
{\sl Kernel Smoothing.}
Chapman \& Hall, London.}

\bye